\documentclass[
 aip, jcp, amsmath,amssymb,
 reprint,%
 groupedaddress,
]{revtex4-1}

\usepackage{rotating}
\usepackage{graphicx}
\usepackage{dcolumn}
\usepackage{bm}
\usepackage{forest}
\usepackage{tikz}
\usepackage{natbib}
\usepackage{siunitx}
\usepackage{CJKutf8}
\usepackage{booktabs}

\usepackage[utf8]{inputenc}
\usepackage[T1]{fontenc}
\usepackage{etoolbox}
\usepackage{color}
\usepackage[normalem]{ulem}

\makeatletter
\def\@email#1#2{%
 \endgroup
 \patchcmd{\titleblock@produce}
  {\frontmatter@RRAPformat}
  {\frontmatter@RRAPformat{\produce@RRAP{#1\href{mailto:#2}{#2}}}\frontmatter@RRAPformat}
  {}{}
}%
\makeatother
\begin{document}

\preprint{AIP/123-QED}

\title[SDVR]{Stochastic Discrete Variable Representation: Tensor Network Eigenstates on General Potential Energy Surfaces}

\author{Hannes Hoppe}
 \email{hannes.hoppe@phys.chem.ethz.ch}
\affiliation{Institute of Molecular Physical Sciences, ETH Zurich, 8093 Zurich, Switzerland
}%

\date{\today}

\begin{abstract}
Tensor network states such as tree tensor network states~(TTNS) or matrix product states~(MPS) allow very compact representations of high-dimensional wave functions.
Using them efficiently relies on a compact representation of the Hamiltonian, and many methods are limited to the sum-of-products~(SOP) form.
General potential energy surfaces~(PESs) do not have this form.
To overcome this issue, a stochastic discrete variable representation~(SDVR) is presented.
The matrix elements of the potential operator in the node-local basis are estimated stochastically by importance sampling of grid configurations.
The estimator of the matrix elements is unbiased, and the PES is only evaluated pointwise.
The variance is reduced by representing parts of the potential with a $2$-mode representation and only sampling the remainder.
Sampling noise biases the eigenvalues. A noise-control scheme mitigates this bias.
Final energies are obtained from an unbiased variational Monte Carlo estimator.
The method is demonstrated by calculating vibrational states of CH$_3$~(6D) and acetonitrile~(12D), as well as the zero-point vibrational energy of uracil~(30D).
For all three systems, the energies agree with the exact references to sub-wavenumber accuracy.
\end{abstract}

\maketitle

\section{Introduction}

Tensor network approaches are a powerful tool in the field of numerical quantum mechanics.
Tree tensor network states~(TTNS)\cite{Shi2006} represent high-dimensional wave functions as tree-shaped tensor contractions.
In the field of chemistry, this is realized in the multiconfiguration time-dependent Hartree~(MCTDH) approach\cite{Meyer1990,Beck2000} and its multi-layer (ML-MCTDH) generalization\cite{Wang2003,Manthe2008}.
A special case of these TTNS is the matrix product state~(MPS), which underlies the density matrix renormalization group~(DMRG) method\cite{White1992,Schollwoeck2011,Baiardi2017,Rakhuba2016,Glaser2023}.

All these methods require matrix elements of the Hamiltonian in the basis of the node-local configurations.
In the case of a sum-of-products~(SOP) Hamiltonian, these matrix elements are cheap to evaluate.
This is the case for most model systems in physics as well as for kinetic energy operators in chemistry.
However, potential energy surfaces~(PESs) typically do not have this form.
There are three main approaches to address this issue.
First, the PES can be refitted into SOP form\cite{Manzhos2006,Avila2015,Ziegler2016}, e.g., by POTFIT\cite{Jaeckle1996} and its successors\cite{Pelaez2013,Schroeder2017,Schroeder2020,Otto2014,Otto2018}. TT-cross\cite{Oseledets2010} and quantics tensor cross interpolation\cite{Ritter2024} are alternatives.
Second, the matrix elements can be computed directly \textit{on-the-fly}, using state-dependent grids.
The correlation discrete variable representation~(CDVR) approach\cite{Manthe1996,Manthe2009,vanHarrevelt2005,Ellerbrock2022,Ellerbrock2023,Ellerbrock2024,Manthe2026,Manthe2017} uses state-dependent quadrature grids for this.
Third, collocation\cite{Wodraszka2018,Wodraszka2020,Wodraszka2021,Simmons2023} requires the Schr\"odinger equation to hold at a set of points.
It needs no matrix elements of the Hamiltonian, and the PES is evaluated exactly at these points.

In this work, a new route is presented, in which the matrix elements are evaluated stochastically.
When applying the discrete variable representation~(DVR) approximation\cite{Light1985,Light2000}, the potential operator is diagonal in the basis of grid functions.
This is combined with the perfect sampling scheme of Ferris and Vidal\cite{Ferris2012}.
It draws independent configurations exactly from a tensor network state~(TNS) without Markov chains.
Here, it is adapted to sample from the individual single-particle functions~(SPFs) and single-hole functions~(SHFs).

Monte Carlo sampling has been used before to optimize states variationally\cite{Sandvik2007,Schuch2008}, especially in the field of neural-network quantum states\cite{Carleo2017}.
It has also been applied to sample adiabatic populations from TNS wave functions by sampling from $\vert\Psi\vert^2$ (Refs.~\onlinecite{worth1998,Cooper2026}).
These approaches, however, estimate the expectation value for a fixed state.
In this work, an operator is sampled in the local configuration space of a node, which then enters the eigenvector equations.
Therefore, sampling from $\vert\Psi\vert^2$ is no longer sufficient.
Unoccupied regions of the Hilbert space also contribute to the matrix elements\cite{Wan2026}.
This issue is addressed by sampling from a mixture distribution\cite{Hesterberg1995,Owen2000}.
Parts of the potential are replaced by an exactly integrable approximation to reduce the variance of the sampled operator.
Diffusion Monte Carlo\cite{McCoy2006} likewise requires only pointwise evaluations of the PES and is well established for vibrational ground states.
It does not, however, yield an explicit representation of the wave function, and excited states require additional approximations such as fixed nodes.

The SDVR method shares some similarities with each of the methods mentioned above.
Similar to the POTFIT family, a simple SOP representation of the dominant features of the PES is generated.
This representation, however, is not a global fit but is adapted to the current state.
Second, the grid points at which the PES is evaluated depend on the current state, as in the CDVR approach.
Finally, as in collocation, the PES is evaluated at specific geometries, but these are used to build matrix representations rather than as collocation points.

Section~\ref{sec:theory} outlines the method.
It is then applied to vibrational states of CH$_3$ and acetonitrile and to the zero-point vibrational energy (ZPVE) of uracil.

\section{Theory}
\label{sec:theory}

In the following, wave functions of $f$ degrees of freedom~(DOFs) are considered.
Each of the DOFs will be represented in a basis of $N_k$ primitive basis functions $\sigma_i^{(k)}$.
These are localized functions, e.g., DVR functions, each associated with a grid point $\xi^{(k)}_i$.
A primitive configuration $\sigma_I=\prod_{k=1}^{f}\sigma^{(k)}_{i_k}$ is labeled by a multi-index $I=(i_1,...,i_f)$ and corresponds to the grid point $\bm{\xi}_I=(\xi^{(1)}_{i_1},...,\xi^{(f)}_{i_f})$.
Any state in the Hilbert space spanned by the configurations $\sigma_I$ can be written as
\begin{equation}
    \vert\Psi\rangle=\sum_I\Psi_I \vert\sigma_I\rangle.
\end{equation}
In practice, this representation is only viable for small systems, because the number of coefficients $\Psi_I$ grows exponentially with $f$.

\subsection{(Tree) Tensor Network States}
\begin{figure*}[t]
    \centering
    \includegraphics{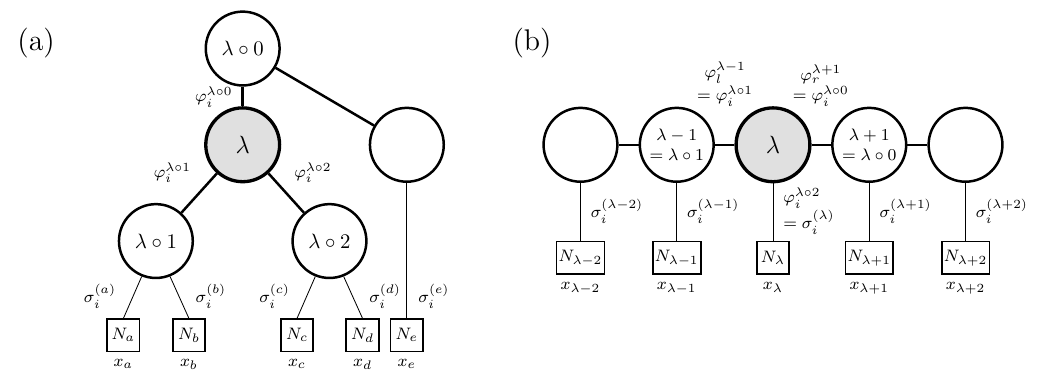}
    \caption[Graphical representation of tree tensor network states.]{Graphical representation of tree tensor network states.
    Circles denote nodes, and the shaded node is the canonical center $(\lambda)$.
    Lines between nodes carry the SPFs and SHFs $\varphi_i^{\lambda\circ\kappa}$.
    Lines ending in squares denote primitive DOFs $x_k$ and carry the primitive basis functions $\sigma_i^{(k)}$.
    The number in the box specifies the primitive basis set size $N_k$.
    Panel (a) shows a general TTNS, panel (b) shows the special case of an MPS.}
    \label{fig:ttns-mps}
\end{figure*}

To circumvent this issue, wave functions can be represented by tensor network states~(TNS).
The specific kind of TNS employed in this work is the tree TNS~(TTNS).
The notation used to specify these TTNS is inspired by the ML-MCTDH notation\cite{Wang2003,Manthe2008,Larsson2024}.
A TTNS is a graph of nodes $(\lambda)$ without loops.
The nodes connected to $(\lambda)$ are denoted $(\lambda\circ\kappa)$, where $(\lambda\circ0)$ corresponds to the parent node and $\kappa=1,...,\nu$ label the child nodes.
A general TTNS can be specified recursively as
\begin{equation}
    \varphi_n^{\lambda}=\sum_{j_1...j_\nu}^{D^{\lambda\circ 1}...D^{\lambda\circ\nu}}A^{\lambda}_{j_1...j_\nu n}\prod_{\kappa=1}^{\nu}\varphi_{j_\kappa}^{\lambda\circ\kappa},
\end{equation}
where $\varphi_n^0=\Psi_n$ is the wave function of the system.
The functions $\varphi_n^{\lambda}$ are called single-particle functions~(SPFs), and $D^{\lambda\circ\kappa}$ is called the \textit{bond dimension}.
At the bottom of the tree, the SPFs are replaced by the primitive functions $\sigma_i^{(k)}$.
An example of a TTNS is given in Fig.~\ref{fig:ttns-mps}(a), which also introduces a graphical notation to simplify specifying a TTNS topology.

In practice, it is useful to work with a TTNS in the \textit{canonical representation}.
In this representation, all nodes except the \textit{canonical center} $(\lambda)$ are isometries, meaning for every node $(\lambda')\neq(\lambda)$
\begin{equation}
    \sum_{J/\eta}A^{\lambda'*}_{J/\eta,k}A^{\lambda'}_{J/\eta,l}=\delta_{kl}.
\end{equation}
Here, $J/\eta$ denotes the superindex $J$ in which the $\eta$th index is omitted, and $J/\eta,k$ adds $k$ as $\eta$th index.
The index $\eta$ denotes the index pointing in the direction of the canonical center $(\lambda)$.
From now on, the tensor at the canonical center will be denoted $\tilde A^{\lambda}$ with a tilde.
With $(\lambda)$ as the canonical center, the functions $\varphi^{\lambda\circ0}_n$ are the normalized single-hole functions (SHFs) of $(\lambda)$, allowing the compact notation
\begin{equation}
    \Phi_J^{\lambda}=\prod_{\kappa=0}^\nu\varphi_{j_\kappa}^{\lambda\circ\kappa},\;\;\vert\Psi\rangle=\sum_J\tilde A^{\lambda}_J\vert\Phi_J^\lambda\rangle.
\end{equation}
Note that $\kappa$ starts at zero here.
The $\Phi_J^{\lambda}$ are called configurations and fulfill $\langle\Phi_J^\lambda\vert\Phi_K^{\lambda}\rangle=\delta_{JK}$.
The canonical center of a TTNS can be shifted to a neighboring node $(\lambda\circ\eta)$ by performing a singular value decomposition~(SVD) of $\tilde A^{\lambda}$ with respect to its index $j_\eta$, where $\tilde A^\lambda=A^{\lambda}\cdot\Sigma\cdot V^\dagger$, $A^\lambda$ becomes an isometry, and $A^{\lambda\circ\eta}\rightarrow \tilde A^{\lambda\circ\eta}$ by contracting with $\Sigma\cdot V^\dagger$ appropriately.
While there are other methods to shift the canonical center, the SVD-based approach is employed in this article, since it leaves the crossed bond in its natural orbital basis, where
\begin{equation}
    \rho_{nm}^{\lambda\circ\kappa}=\sum_{J/\kappa} \tilde A^{\lambda*}_{J/\kappa,n}\tilde A_{J/\kappa,m}^{\lambda}
\end{equation}
is diagonal.
This allows for the interpretation of $\rho_{nn}^{\lambda\circ\kappa}$ as the population of $\varphi^{\lambda\circ\kappa}_{n}$.

An MPS is a special case of a TTNS where each node $(\lambda)$ has exactly two neighboring nodes and carries a primitive DOF $x_\lambda$, as shown in Fig.~\ref{fig:ttns-mps}(b).
The only exceptions are the first and last nodes, which have only one neighboring node.
For an MPS, a node and its neighbors are labeled $(\lambda-1),(\lambda),(\lambda+1)$.
These labels replace $(\lambda\circ1),(\lambda),(\lambda\circ 0)$.
Also note that conventionally the node labels in an MPS start at 1.

\subsubsection{Eigenstate Calculation}
TTNS provide an efficient framework for eigenstate calculations of quantum systems\cite{Larsson2025,Larsson2026}.
Eigenstates are obtained by sweeps according to the scheme presented by Larsson\cite{Larsson2019}.
In this scheme, an Euler tour is performed along the tree, and each node is visited and made the canonical center.
Then, all tensors except $\tilde A^{\lambda}$ are frozen, and
\begin{equation}
    \min_{\tilde A^{\lambda}}\frac{\langle\Psi\vert\hat H\vert\Psi\rangle}{\langle\Psi\vert\Psi\rangle}
\end{equation}
is solved.
Here, $\hat H$ denotes the Hamiltonian of the system.
This is done by solving the local eigenproblem
\begin{equation}
    \sum_{J}\langle\Phi_K^{\lambda}\vert\hat H\vert\Phi_J^{\lambda}\rangle \tilde A^{\lambda}_{J}=E \tilde A^{\lambda}_K.
\end{equation}
The lowest eigenvector then replaces $\tilde A^{\lambda}$, and the canonical center is shifted to the next node.
This is repeated until convergence.

Excited states can be obtained by updating
\begin{equation}
    \hat H\rightarrow \hat H+\Delta\sum_{n}\vert\Psi_n\rangle\langle\Psi_n\vert
    \label{eq:penalty}
\end{equation}
and thus shifting all previously obtained states $\Psi_n$ up in energy.
Alternatively, state-averaged solvers as given in Refs.~\onlinecite{Larsson2019,Manthe2008SA,Hoppe2024} can be used, in which several states share all tensors except the one at the canonical center.
When the canonical center is moved, the reduced density matrices of these states are averaged.

The eigenstate calculation relies on matrix elements of the Hamiltonian.
Obtaining the matrix elements of the potential is the main subject of this article.

\subsection{Form of the Hamiltonian}

Most approaches employing TTNS as wave function ansatz rely on a Hamiltonian in sum-of-products~(SOP) form
\begin{equation}
    \hat H_\text{SOP}=\sum_{i}c_i\prod_{k=1}^{f}\hat h_{k}^{i}.
\end{equation}
Each primitive operator $\hat h_{k}^{i}$ acts on a single coordinate $x_{k}$.
The matrix elements of $\hat H_\text{SOP}$ therefore factorize into those of the primitive operators, which the usual recursive scheme\cite{Beck2000,Manthe2008} evaluates.

In this work, a more general form of $\hat H$ will be investigated:
\begin{equation}
    \hat H=\hat H_\text{SOP}+\hat V.\label{eq:fullH}
\end{equation}
The operator $\hat V$ is defined by mapping a grid point $\bm{\xi}_I$ to a value $V(\bm{\xi}_I)$.
Let $\mathcal{F}$ denote the set of all configurations $\sigma_I$ present in the Hilbert space.
In the DVR approximation, this operator is diagonal and given by
\begin{equation}
    \hat V_\text{DVR}=\sum_{I\in\mathcal{F}}V(\bm{\xi}_I)\vert\sigma_I\rangle\langle\sigma_I\vert.\label{eq:exactDVR}
\end{equation}
In practice, this operator cannot be used for most systems, since the sum runs over all $\vert\mathcal{F}\vert=\prod_{k=1}^{f} N_k$ configurations\cite{Avila2009}.
The SOP part of $\hat H$ will still be evaluated using the established recursive scheme, and only the matrix elements of $\hat V$ will be treated differently.

\subsection{Stochastic Operator}

\subsubsection{Importance-Sampled Operator}

Given the canonical center at $(\lambda)$, the matrix elements of the operator defined in Eq.~\eqref{eq:exactDVR} are given by
\begin{multline}
    \langle\Phi_J^{\lambda}\vert\hat V_\text{DVR}\vert\Phi_K^\lambda\rangle=\sum_{I\in\mathcal{F}} V(\bm{\xi}_I)\langle\Phi_J^\lambda\vert\sigma_I\rangle\langle\sigma_I\vert\Phi_K^\lambda\rangle\\
    =\sum_{I\in\mathcal{F}}V(\bm{\xi}_I)\prod_{\kappa=0}^\nu\varphi_{j_\kappa}^{\lambda\circ\kappa*}(I_\kappa)\varphi_{k_\kappa}^{\lambda\circ\kappa}(I_\kappa)\label{eq:matrixelements}
\end{multline}
where $I_\kappa$ denotes the part of the configuration $I=(I_1,...,I_\nu,I_0)$ that belongs to the subtree of $\varphi_n^{\lambda\circ\kappa}$.
Note that in the MPS case $\varphi^{\lambda\circ2}_n(i_\lambda)=\sigma^{(\lambda)}_n(i_\lambda)=\delta_{ni_\lambda}$.
In SDVR, this sum over all configurations is replaced by a sum over a stochastically chosen multiset of configurations.
Let $\mathcal{S}$ be a multiset of independent draws, where configuration $I$ is drawn with probability
\begin{equation}
    P(I)= \prod_{\kappa=0}^{\nu}P^{\lambda\circ\kappa}(I_\kappa).
\end{equation}
Because $P$ factorizes, the subtrees $(\lambda\circ\kappa)$ can be sampled independently.
The details of this probability distribution are given in Sec.~\ref{sec:sampling-distribution}.
This sampling-based approach yields the most general form of the SDVR operator
\begin{equation}
     \langle\Phi_J^{\lambda}\vert\hat V_\text{SDVR}\vert\Phi_K^\lambda\rangle=\frac{1}{|\mathcal{S}|}\sum_{I\in\mathcal{S}} \frac{V(\bm{\xi}_I)}{P(I)}\prod_{\kappa=0}^\nu\varphi_{j_\kappa}^{\lambda\circ\kappa*}(I_\kappa)\varphi_{k_\kappa}^{\lambda\circ\kappa}(I_\kappa).\label{eq:foundeq}
\end{equation}
$\vert\mathcal{S}\vert$ denotes the number of drawn samples.
Equation~\eqref{eq:foundeq} is an unbiased estimator of Eq.~\eqref{eq:matrixelements} if $P(I)\neq 0$ wherever the product of amplitudes is nonzero.
This condition is always fulfilled for the distribution of Sec.~\ref{sec:sampling-distribution}, since $\mu>0$ ensures $\pi_n^{\lambda\circ\kappa}>0$ for every $n$.
Since the same multiset $\mathcal{S}$ is used for all pairs $(J,K)$, the estimated matrix is exactly Hermitian for real $V$.

\subsubsection{Sampling Distribution}
\label{sec:sampling-distribution}
For an expectation value $\langle\Psi\vert\hat O\vert\Psi\rangle$, the measure $P(I)=\vert \Psi(I)|^2$ is the natural measure and gives bounded weights\cite{Cooper2026}.
Sampling the matrix representation of an operator is more involved.
Unoccupied SPFs and SHFs are important for the representation of the potential operator, since they appear in Eq.~\eqref{eq:foundeq}.
Sampling according to $|\Psi|^2$ would entirely neglect these\cite{Wan2026}.
Therefore, a mixture distribution is employed\cite{Hesterberg1995,Owen2000}
\begin{equation}
    \pi^{\lambda\circ\kappa}_n=(1-\mu)\frac{\rho_{nn}^{\lambda\circ\kappa}}{\text{tr}\rho^{\lambda\circ\kappa}}+\frac{\mu}{D^{\lambda\circ\kappa}}, \quad\mu\in(0,1],
\end{equation}
defined using the natural populations $\rho_{nn}^{\lambda\circ\kappa}$ of the bond and the bond dimension $D^{\lambda\circ\kappa}$.
At $\mu=1$, the SPFs and SHFs are sampled uniformly.
As $\mu\rightarrow0$, each subtree approaches its marginal of $|\Psi|^2$, exactly so if $\rho^{\lambda\circ\kappa}$ is diagonal. 
Starting at the canonical center $(\lambda)$, a random SPF or SHF is chosen according to the probability distribution defined by $\pi^{\lambda\circ\kappa}_n$.
This is done for each edge independently.
Then, a primitive configuration $\vert\sigma_{I_\kappa}\rangle$ is selected from the subtree corresponding to node $(\lambda\circ\kappa)$.
This is done with probability $\vert \varphi_n^{\lambda\circ\kappa}(I_\kappa)\vert^2$.
The exact procedure for this is explained in Sec.~\ref{sec:perfect-sampling}.
The resulting marginal probability of drawing $I_\kappa$ is
\begin{equation}
    P^{\lambda\circ\kappa}(I_\kappa)=\sum_n \pi^{\lambda\circ\kappa}_n \vert \varphi_n^{\lambda\circ\kappa}(I_\kappa)\vert^2,\label{eq:marginal}
\end{equation}
which is normalized, since both $\varphi_n^{\lambda\circ\kappa}$ and $\pi^{\lambda\circ\kappa}_n$ are normalized.
If $(\lambda)$ carries one or more primitive DOFs $x_k$, all grid points of these DOFs are taken into account for each sample of the environment.
The probability function in Eq.~\eqref{eq:foundeq} is then replaced with the probability of the environment,
\begin{equation}
    P_\text{env}(I)=\prod_{\kappa\text{ not primitive}} P^{\lambda\circ\kappa}(I_\kappa).
\end{equation}
This is a form of Rao--Blackwellization\cite{Casella1996}, and it is employed to circumvent the problem of unoccupied primitive grid configurations.
In this case, $\vert\mathcal{S}\vert$ denotes the number of PES evaluations per node update, and $\vert\mathcal{S}\vert/N_\lambda$ environments are drawn, where $N_\lambda$ is the number of local primitive grid points of $(\lambda)$.

The sampled operator is bounded, since
\begin{equation}
    \frac{\vert\varphi^{\lambda\circ\kappa}_n(I_\kappa)\varphi^{\lambda\circ\kappa}_m(I_\kappa)\vert}{P^{\lambda\circ\kappa}(I_\kappa)}\leq \frac{1+\delta_{nm}}{2\sqrt{\pi_n^{\lambda\circ\kappa}\pi_m^{\lambda\circ\kappa}}}\leq \frac{D^{\lambda\circ\kappa}}{\mu}.\label{eq:finiteop}
\end{equation}
The first inequality follows from $\pi_n|\varphi_n|^2+\pi_m|\varphi_m|^2\geq2\sqrt{\pi_n\pi_m}|\varphi_n\varphi_m|$, and the second from $\pi_n^{\lambda\circ\kappa}\geq\mu/D^{\lambda\circ\kappa}$. 
The variance $\mathrm{Var}$ of a matrix element of Eq.~\eqref{eq:foundeq} is
\begin{multline}
    \mathrm{Var} = \frac{1}{\vert\mathcal{S}\vert}\Bigg[\sum_{I\in\mathcal{F}}\frac{\vert V(\bm{\xi}_I)\vert^2\vert \prod_{\kappa}\varphi^{\lambda\circ\kappa}_{j_\kappa}(I_\kappa)\varphi^{\lambda\circ\kappa}_{k_\kappa}(I_\kappa)\vert^2}{P(I)}\\
    -\vert\langle\Phi_J^\lambda\vert\hat V_\text{DVR}\vert\Phi_K^\lambda\rangle\vert^2\Bigg].
\end{multline}
Given Eq.~\eqref{eq:finiteop} and the Cauchy--Schwarz inequality $\sum_I \vert \prod_{\kappa}\varphi^{\lambda\circ\kappa}_{j_\kappa}(I_\kappa)\varphi^{\lambda\circ\kappa}_{k_\kappa}(I_\kappa)\vert\leq 1$, the variance of a matrix element is bounded by
\begin{equation}
    \mathrm{Var} \leq \frac{1}{\vert\mathcal{S}\vert}\prod_{\kappa\text{ not primitive}}\frac{D^{\lambda\circ\kappa}}{\mu}\max_{I}\vert V(\bm{\xi}_I)\vert^2.\label{eq:varbound}
\end{equation}
This means that, targeting a fixed accuracy of the sampled matrix elements, it is sufficient to increase the number of samples proportionally to the product of the bond dimensions of a node.
In the case of an MPS, this means a quadratic increase of the number of samples with the bond dimension, since the primitive DOF of each node is summed exactly.
The error of the eigenvalues depends on the noise of the whole matrix and can grow with its dimension.
The bound is not necessarily tight, and the actual variance can be estimated from the same samples without additional PES evaluations.
This also shows that the variance depends quadratically on the value of $V$ at the grid points.
A scheme to reduce this will be proposed in Sec.~\ref{sec:fitting}.
Since the sampling distribution depends on SPFs and SHFs, a new multiset $\mathcal{S}$ needs to be redrawn whenever the canonical center changes.

\subsubsection{Perfect Sampling}
\label{sec:perfect-sampling}

The sampling scheme\cite{Ferris2012,White2009,Stoudenmire2010} is first outlined for an MPS.
Assume that a configuration for the DOFs $x_1,...,x_{\lambda-1}$ needs to be drawn according to $\vert\varphi^{\lambda-1}_l(I_<)\vert^2$.

\begin{enumerate}
    \item First, starting at $(\lambda)$, choose $\varphi_l^{\lambda-1}$ with probability $\pi^{\lambda-1}_l$. Set $\mathbf{w}=\mathbf{e}_l=(0,0,...,1,...0)$ and $k=\lambda-1$.
    \item Contract $\mathbf{w}$ into $A^{k}$, yielding $B_{m{i_k}}=\sum_{n}A_{m{i_k}n}w_n$.
    \item Draw a grid point $i_k$ according to the distribution $P(i_k\vert i_{k+1}...i_{\lambda-1})=\sum_m |B_{mi_k}|^2/\sum_m|w_m|^2$.
    \item Set $w_n=B_{ni_k}$ and $k = k - 1$, and repeat from step 2 as long as $k\geq1$.
\end{enumerate}

Due to $A^{k}$ being an isometry, the probabilities of each step sum to 1.
This procedure is repeated for the right environment of $(\lambda)$.

In a TTNS, a neighboring environment might itself branch. 
The configuration of the drawn SPF $\varphi^{\lambda\circ\kappa}_n$ follows exactly from the perfect sampling scheme of Ref.~\onlinecite{Ferris2012}. Contracting the subtree at this configuration gives the amplitudes of all SPFs of $(\lambda\circ\kappa)$.
From the perspective of $(\lambda)$, the SHFs play the same role as the SPFs, and the environment behind them is thus sampled like any other subtree.

Equation~\eqref{eq:foundeq} requires the evaluation of $\varphi_{j_\kappa}^{\lambda\circ\kappa}(I_\kappa)$ for all $j_\kappa$, not only for the selected SPF.
After $I_<$ has been drawn, these amplitudes are obtained by a second contraction from the first site towards $(\lambda)$, at a cost of $\mathcal{O}(D^2)$ per site.
These amplitudes enter both the importance weight $1/P(I)$, through Eq.~\eqref{eq:marginal}, and the matrix elements of Eq.~\eqref{eq:foundeq}.
For an MPS, one sample costs about $\mathcal{O}(fND^2)$ operations. Here, $N$ and $D$ are the largest primitive grid size and bond dimension.

\subsubsection{Applying the Operator}
The algorithm used for eigenstate calculation does not require a matrix representation of $\hat H$, only the action of the operator is required.
Thus, it can be applied \textit{on-the-fly}.
With $g_{ml}=\varphi_{l}^{\lambda-1}(I_{m,<})$, $h_{mr}=\varphi_{r}^{\lambda+1}(I_{m,>})$, and $W_{mi}=V(\bm{\xi}_{m,<},\xi_{i}^{(\lambda)},\bm{\xi}_{m,>})/(|\mathcal{S}|/N_\lambda\,P_\text{env}(I_m))$,
\begin{equation}
    \left(\hat V_\text{SDVR}\tilde A^{\lambda}\right)_{lir}=\sum_{m=1}^{\vert\mathcal{S}\vert/N_\lambda}g_{ml}^*h^*_{mr}W_{mi}\sum_{l'r'}g_{ml'}\tilde A_{l'ir'}h_{mr'}.
\end{equation}

This operator application costs $\mathcal{O}(\vert\mathcal{S}\vert D^{\lambda-1} D^{\lambda+1})$ operations.
Since $\hat V_\text{SDVR}$ is, for each $i$, a sum of $|\mathcal{S}|/N_\lambda$ rank-one contributions, at least $N_\lambda D^{\lambda-1}D^{\lambda+1}$ PES evaluations are required for a full-rank representation.
For a TTNS, this expression generalizes with one environment factor per edge of $(\lambda)$.

\subsection{Variance Reduction by an $n$-Mode Control Variate}
\label{sec:fitting}

As shown in Sec.~\ref{sec:sampling-distribution}, the variance of the sampled operator is bounded only through the largest magnitude of $V$ on the grid, which can be large.
This can be mitigated by splitting $V = F + V_r$, where $F$ is a cheap $n$-mode representation of the PES:
\begin{equation}
    F(\mathbf{x})=v_0+\sum_k f_k(x_k)+\sum_{a<b}f_{ab}(x_a,x_b)+...
\end{equation}
Here, $V(x_k;\bm{\xi}_\text{ref})$ denotes the potential with all coordinates except $x_k$ fixed at the reference configuration $\bm{\xi}_\text{ref}$. Then $v_0=V(\bm{\xi}_\text{ref})$, $f_k(x_k)=V(x_k;\bm{\xi}_\text{ref})-v_0$, and $f_{ab}(x_a,x_b)=V(x_a,x_b;\bm{\xi}_\text{ref})-f_a(x_a)-f_b(x_b)-v_0$.
$F$ is easily obtained and can be included exactly in $\hat H_\text{SOP}$, and only the remainder $V_r$ needs to be sampled, which reduces the variance if $\vert V_r\vert$ is small compared with $\vert V\vert$.
Here, $F$ acts as a control variate\cite{Owen2000}.
To include $F$ in $\hat H_\text{SOP}$, each $2$-mode term $f_{ab}$, given on the $N_a\times N_b$ grid, is decomposed by an SVD into a sum of products of $1$-mode functions.
This expansion can be truncated arbitrarily, since the truncation error becomes part of the remainder $V_r$ and is thus sampled.
To define an $n$-mode representation, a reference configuration needs to be defined.
In test calculations, the expectation value $\langle x_k\rangle$ and the most populated grid point of each mode made no systematic difference.
This work uses the latter.
The reference configuration is determined from the current state and is updated after each node update.

In principle, $F$ is not limited to an $n$-mode representation.
To further reduce the variance, any $F$ with exactly computable matrix elements can serve.
It must not depend on the samples used for $V_r$.
In this sense, SDVR can be regarded as an unbiased stochastic correction to any SOP fit of the PES.

\subsection{Noise Control for Eigenvectors}
\label{sec:noise-control}
While the sampled operator is unbiased, its eigenvalues are not\cite{Lim2017,Blunt2018}.
The lowest eigenvalue of the sampled operator is biased downward on average compared to the correct eigenvalue.
This follows from Jensen's inequality, since the lowest eigenvalue is a concave function of the matrix.
The eigenvector is biased as well, since it adapts to the noise, so that the exact energy of the optimized state lies above the lowest energy attainable with the ansatz.
To mitigate this issue, a simple ad-hoc scheme is proposed for the eigensolver.
The eigenvalue iteration starts with $\tilde A^\lambda_\text{old}$, and then the node-local eigenvalue problem is solved, yielding $\tilde A^\lambda_\text{new}$.
Note that the SDVR operator is frozen during the iteration, the same samples are used for every operator application.
Next, an independent multiset (of equal size) is drawn.
With it, $\hat H$ is evaluated in the span of the old and the new $\tilde A$.
Then, using this representation of $\hat H$, the energy $E(\tau)$ of the normalized state $\tilde A_\tau=(1-\tau)\tilde A^\lambda_\text{old}+\tau \tilde A^{\lambda}_\text{new}$ is minimized for $\tau\in[0,1]$ by solving a quadratic equation.
If $\tau<0.1$ or $\tau>0.9$, $\tau$ is set to 0 or 1, respectively, to further reduce noise.
The step is accepted and $\tilde A_\tau$ becomes the new $\tilde A^\lambda$ iff $E(\tau)<E(0)$, otherwise $\tilde A^\lambda$ is not touched. This resembles the linear method of variational Monte Carlo\cite{Toulouse2007}.

Final energies come from the variational Monte Carlo~(VMC) estimator
\begin{equation}
    \langle E\rangle=\frac{1}{\vert\mathcal{M}\vert}\sum_{I\in\mathcal{M}}\frac{\langle\sigma_I\vert \hat H\vert\Psi\rangle}{\langle\sigma_I\vert \Psi\rangle},
    \label{eq:vmc}
\end{equation}
where $\mathcal{M}$ is a multiset of independent draws from $\vert\Psi\vert^2$.
For $\hat V_\text{DVR}$, the numerator is $V(\bm{\xi}_I)\Psi(I)$. 
For $\hat H_\text{SOP}$, it is obtained by contracting the TTNS with the primitive operators along the sampled configuration.
The statistical error of $\langle E\rangle$ is estimated as $\sqrt{\mathrm{Var}(E_L)/\vert\mathcal{M}\vert}$ from the local energies $E_L(I)=\langle\sigma_I\vert \hat H\vert\Psi\rangle/\langle\sigma_I\vert \Psi\rangle$.
This estimator is unbiased.
Its mean is an upper bound to the ground-state eigenvalue of the full-grid DVR Hamiltonian, Eq.~\eqref{eq:fullH}.
Individual estimates can fall below this bound due to statistical noise.
Equation~\eqref{eq:vmc} employs $\hat H$ without the penalty term of Eq.~\eqref{eq:penalty}.
For excited states, $\langle E\rangle$ carries no variational guarantee.
Its error bars can be unreliable for heavy-tailed local energies near nodes\cite{Trail2008,Wan2026}.

\section{Results and Discussion}
The ground state is initialized as a Hartree product, and each excited state is initialized as a slightly perturbed copy of the previously computed state.
In all calculations, the local eigenproblem is solved with a standard Lanczos solver until $\tilde A^{\lambda}$ changes by less than $10^{-8}$, and the sweeps are considered converged once the VMC energy $\langle E\rangle$ (cp.\ Eq.~\eqref{eq:vmc}) changes by less than $4\times10^{-7}$~Hartree ($\approx$\SI{0.09}{cm^{-1}}) between two sweeps.
The statistical error of $\langle E\rangle$ is monitored alongside.
In all calculations, the SVD of the $2$-mode terms of the control variate (Sec.~\ref{sec:fitting}) is truncated at a relative threshold of $10^{-6}$.
All excited states are obtained with the penalty approach of Eq.~\eqref{eq:penalty} and $\Delta=0.1$~Hartree ($\approx\SI{2.2e4}{cm^{-1}}$).
In this approach, the errors of all lower states enter the penalty, so that errors accumulate for higher states, and their energies are not bounded from below by the exact values.
Both the convergence check and the final energies use VMC estimates with $\vert\mathcal{M}\vert=10^6$.
For CH$_3$ and acetonitrile, their standard errors are below \SI{0.01}{cm^{-1}}.
Throughout, \textit{exact} refers to calculations with the same TTNS in which the potential matrix elements are evaluated without sampling, either with the SOP form of the Hamiltonian or by summing over the whole grid.
Deviations $\Delta E=E-E_\text{exact}$ are signed differences and can have either sign.
\subsection{Methyl Radical}

\begin{figure}
    \centering
    \includegraphics{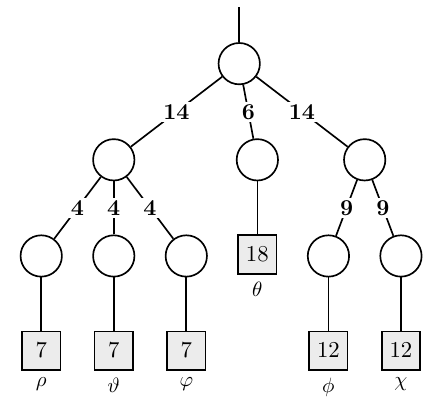}
    \caption{TTNS topology employed in the calculation of the vibrational eigenstates of CH$_3$.}
\label{fig:ch3-tree}
\end{figure}

Vibrational eigenstates of the methyl radical~(6D) are computed with the PES of Medvedev~\textit{et al.}\cite{Medvedev2006}.
The kinetic energy operator and the potential-optimized DVR basis are those of Ref.~\onlinecite{evenhuis2007}.
The TTNS topology, bond dimensions, and primitive grid sizes are given in Fig.~\ref{fig:ch3-tree}.

Table~\ref{tab:ch3_vmc} lists the energies obtained for different combinations of $\mu$ and $\vert\mathcal{S}\vert$.
The exact energies are the most accurate energies attainable with the given TTNS topology and the full-grid DVR potential of Eq.~\eqref{eq:exactDVR}.
The results for $\vert\mathcal{S}\vert=0$ employ only the $2$-mode representation of the PES.
This is already a good approximation for many states: about a third of the excitation energies deviate from the exact result by less than \SI{1}{cm^{-1}}.
Still, the mean absolute error~(MAE) is about \SI{5}{cm^{-1}}, and the excitation energy of state 26 is overestimated by more than \SI{75}{cm^{-1}}.
Sampling the remainder of the potential removes most of this error.
With $\vert\mathcal{S}\vert=10^4$, the MAE drops to \SIrange{0.2}{0.3}{cm^{-1}}, and the largest deviations are about \SI{1}{cm^{-1}}.
For state 26, the deviation is reduced to less than \SI{0.2}{cm^{-1}}, which shows the importance of the sampled correction.
The full grid has about $9\times10^5$ points, and one set of $10^4$ samples is small in comparison.
Increasing the number of samples to $10^5$ reduces the MAE by almost an order of magnitude, to \SI{0.03}{cm^{-1}}.
At $\vert\mathcal{S}\vert=10^4$, the MAE is \SI{0.21}{cm^{-1}} for $\mu=0.25$ and \SI{0.30}{cm^{-1}} for $\mu=0.75$.
At $10^5$, both give \SI{0.03}{cm^{-1}}.
At $\vert\mathcal{S}\vert=10^4$, the ZPVE and most excitation energies are too high: 26 of 35 for $\mu=0.25$ and 29 of 35 for $\mu=0.75$.
The VMC estimator is unbiased, and its standard error is below \SI{0.01}{cm^{-1}}.
The overestimation therefore reflects the eigenvector bias of Sec.~\ref{sec:noise-control}, which decreases with the number of samples.
While each node update is stochastic, a calculation comprises many independent node updates over several sweeps, whose fluctuations average out. The final energies therefore vary little between independent runs.

\begin{table*}
\centering
\caption[Vibrational energies of the methyl radical.]{Zero-point energy and vibrational excitation energies of the methyl radical in \si{cm^{-1}}.
The exact results are obtained by using the full-grid DVR potential (sampling over all configurations exactly once with uniform weight).
All other results are obtained with the given combination of $\mu$ and $\vert\mathcal{S}\vert$.
$\vert \mathcal{S}\vert=0$ corresponds to only employing a $2$-mode representation, but no additional sampling.
Differences $\Delta E = E - E_\text{exact}$ indicate the deviation from the exact result.
State 1 is the ZPVE; all other energies are excitation energies relative to the ground state.
MAE and RMS are computed over the excitation energies only.}
\label{tab:ch3_vmc}
\footnotesize
\begin{tabular}{rr|rr||rr|rr||rr|rr}
\toprule
 & exact & \multicolumn{2}{c}{\begin{tabular}{@{}c@{}}$\vert\mathcal{S}\vert=0$\end{tabular}} & \multicolumn{2}{c}{\begin{tabular}{@{}c@{}}$\mu = 0.25$ \\ $|\mathcal{S}| = 10^4$\end{tabular}} & \multicolumn{2}{c}{\begin{tabular}{@{}c@{}}$\mu = 0.25$ \\ $|\mathcal{S}| = 10^5$\end{tabular}} & \multicolumn{2}{c}{\begin{tabular}{@{}c@{}}$\mu = 0.75$ \\ $|\mathcal{S}| = 10^4$\end{tabular}} & \multicolumn{2}{c}{\begin{tabular}{@{}c@{}}$\mu = 0.75$ \\ $|\mathcal{S}| = 10^5$\end{tabular}} \\\hline
\cmidrule(lr){3-4} \cmidrule(lr){5-6} \cmidrule(lr){7-8} \cmidrule(lr){9-10} \cmidrule(lr){11-12}
State & $E$ & $E$ & $\Delta E$ & $E$ & $\Delta E$ & $E$ & $\Delta E$ & $E$ & $\Delta E$ & $E$ & $\Delta E$ \\
\midrule
1 & 6445.20 & 6446.09 & 0.89 & 6445.40 & 0.20 & 6445.22 & 0.02 & 6445.25 & 0.05 & 6445.22 & 0.02 \\
\midrule
2 & 591.47 & 591.83 & 0.36 & 592.56 & 1.09 & 591.44 & -0.03 & 591.42 & -0.05 & 591.47 & 0.00 \\
3 & 1265.70 & 1265.86 & 0.16 & 1265.68 & -0.02 & 1265.69 & -0.01 & 1265.79 & 0.09 & 1265.68 & -0.02 \\
4 & 1388.21 & 1388.74 & 0.53 & 1388.15 & -0.06 & 1388.30 & 0.09 & 1388.46 & 0.25 & 1388.18 & -0.03 \\
5 & 1388.44 & 1390.28 & 1.84 & 1388.25 & -0.19 & 1388.37 & -0.07 & 1388.71 & 0.27 & 1388.41 & -0.03 \\
6 & 1991.37 & 1992.35 & 0.98 & 1991.42 & 0.05 & 1991.44 & 0.07 & 1991.67 & 0.30 & 1991.42 & 0.05 \\\hline
7 & 1991.92 & 1993.98 & 2.06 & 1992.14 & 0.22 & 1991.80 & -0.12 & 1991.70 & -0.22 & 1991.90 & -0.02 \\
8 & 1993.40 & 1994.01 & 0.61 & 1993.29 & -0.11 & 1993.38 & -0.02 & 1993.51 & 0.11 & 1993.36 & -0.04 \\
9 & 2673.26 & 2674.21 & 0.95 & 2673.42 & 0.16 & 2673.29 & 0.03 & 2673.66 & 0.40 & 2673.28 & 0.02 \\
10 & 2674.19 & 2679.50 & 5.31 & 2674.46 & 0.27 & 2674.19 & 0.00 & 2674.26 & 0.07 & 2674.14 & -0.05 \\
11 & 2750.68 & 2755.67 & 4.99 & 2750.79 & 0.11 & 2750.63 & -0.05 & 2751.27 & 0.59 & 2750.67 & -0.01 \\
12 & 2762.28 & 2765.72 & 3.44 & 2762.24 & -0.04 & 2762.28 & 0.00 & 2762.33 & 0.05 & 2762.23 & -0.05 \\\hline
13 & 2767.57 & 2767.98 & 0.41 & 2767.58 & 0.01 & 2767.56 & -0.01 & 2767.53 & -0.04 & 2767.53 & -0.04 \\
14 & 2767.60 & 2769.13 & 1.53 & 2767.86 & 0.26 & 2767.64 & 0.04 & 2767.65 & 0.05 & 2767.63 & 0.03 \\
15 & 2989.75 & 2990.70 & 0.95 & 2989.92 & 0.17 & 2989.75 & 0.00 & 2990.20 & 0.45 & 2989.74 & -0.01 \\
16 & 3142.39 & 3143.06 & 0.67 & 3142.39 & 0.00 & 3142.43 & 0.04 & 3142.50 & 0.11 & 3142.38 & -0.01 \\
17 & 3142.58 & 3143.71 & 1.13 & 3142.62 & 0.04 & 3142.49 & -0.09 & 3142.88 & 0.30 & 3142.53 & -0.05 \\
18 & 3367.32 & 3373.20 & 5.88 & 3368.13 & 0.81 & 3367.33 & 0.01 & 3368.05 & 0.73 & 3367.33 & 0.01 \\\hline
19 & 3382.71 & 3385.75 & 3.04 & 3382.63 & -0.08 & 3382.70 & -0.01 & 3382.76 & 0.05 & 3382.71 & 0.00 \\
20 & 3382.80 & 3387.08 & 4.28 & 3382.86 & 0.06 & 3382.78 & -0.02 & 3383.06 & 0.26 & 3382.80 & 0.00 \\
21 & 3406.55 & 3407.61 & 1.06 & 3406.54 & -0.01 & 3406.54 & -0.01 & 3407.00 & 0.45 & 3406.52 & -0.03 \\
22 & 3407.85 & 3408.55 & 0.70 & 3408.14 & 0.29 & 3407.79 & -0.06 & 3408.00 & 0.15 & 3407.81 & -0.04 \\
23 & 3556.76 & 3557.56 & 0.80 & 3556.92 & 0.16 & 3556.75 & -0.01 & 3556.98 & 0.22 & 3556.71 & -0.05 \\
24 & 3573.12 & 3574.99 & 1.87 & 3573.41 & 0.29 & 3573.09 & -0.03 & 3573.35 & 0.23 & 3573.07 & -0.05 \\\hline
25 & 3709.36 & 3713.67 & 4.31 & 3709.37 & 0.01 & 3709.39 & 0.03 & 3709.56 & 0.20 & 3709.38 & 0.02 \\
26 & 3709.60 & 3786.70 & 77.10 & 3709.43 & -0.17 & 3709.60 & 0.00 & 3709.57 & -0.03 & 3709.50 & -0.10 \\
27 & 4057.81 & 4064.00 & 6.19 & 4058.37 & 0.56 & 4057.79 & -0.02 & 4058.65 & 0.84 & 4057.79 & -0.02 \\
28 & 4072.72 & 4075.65 & 2.93 & 4072.98 & 0.26 & 4072.77 & 0.05 & 4072.86 & 0.14 & 4072.72 & 0.00 \\
29 & 4072.95 & 4076.07 & 3.12 & 4073.24 & 0.29 & 4072.91 & -0.04 & 4073.45 & 0.50 & 4072.92 & -0.03 \\
30 & 4107.79 & 4113.70 & 5.91 & 4108.13 & 0.34 & 4107.80 & 0.01 & 4108.74 & 0.95 & 4107.81 & 0.02 \\\hline
31 & 4108.48 & 4124.50 & 16.02 & 4108.81 & 0.33 & 4108.47 & -0.01 & 4109.48 & 1.00 & 4108.43 & -0.05 \\
32 & 4137.47 & 4140.21 & 2.74 & 4137.56 & 0.09 & 4137.50 & 0.03 & 4137.49 & 0.02 & 4137.53 & 0.06 \\
33 & 4138.23 & 4140.81 & 2.58 & 4138.54 & 0.31 & 4138.24 & 0.01 & 4138.21 & -0.02 & 4138.20 & -0.03 \\
34 & 4178.36 & 4179.38 & 1.02 & 4178.70 & 0.34 & 4178.32 & -0.04 & 4179.14 & 0.78 & 4178.37 & 0.01 \\
35 & 4179.92 & 4182.50 & 2.58 & 4180.11 & 0.19 & 4179.91 & -0.01 & 4179.66 & -0.26 & 4179.86 & -0.06 \\
36 & 4234.73 & 4238.29 & 3.56 & 4234.85 & 0.12 & 4234.76 & 0.03 & 4234.92 & 0.19 & 4234.74 & 0.01 \\\hline\hline
\midrule
MAE &  &  & 4.90 &  & 0.21 &  & 0.03 &  & 0.30 &  & 0.03 \\
RMS &  &  & 13.62 &  & 0.31 &  & 0.04 &  & 0.40 &  & 0.04 \\
\bottomrule
\end{tabular}
\end{table*}

This example shows that increasing the number of samples systematically improves the results.
Compared to the exact reference, about three times as many sweeps were necessary to converge, independent of the sample size.
For the larger sample size, the PES is therefore evaluated more often in total than the grid has points.
This is a consequence of the compact, potential-optimized primitive basis, which makes the full grid small.

For comparison, the same system is treated with the non-hierarchical CDVR~(nhCDVR) approach\cite{Ellerbrock2022,Ellerbrock2024}.
In the nhCDVR, the potential operator is represented on state-dependent quadrature grids, which are combined by a nested Smolyak construction\cite{Smolyak1963,Avila2009}.
Owing to this construction, the effective Hamiltonian depends on the tree topology, and the tree has to be designed carefully.
Moreover, correlation between coordinates that are far apart in the tree can be treated less accurately than correlation between neighboring coordinates.
On the other hand, the nhCDVR is very point-efficient.
For this tree, it needs only about $2.5\times10^3$ points per node update.
With the tree of Fig.~\ref{fig:ch3-tree}, the nhCDVR gives a ZPVE of \SI{6448.16}{cm^{-1}}, \SI{3.0}{cm^{-1}} above the exact value of \SI{6445.20}{cm^{-1}} in Table~\ref{tab:ch3_vmc}.
Its effective Hamiltonian is only approximate for this tree.
Per node update, SDVR thus requires about 4 and 40 times as many PES evaluations for $\vert\mathcal{S}\vert=10^4$ and $10^5$, respectively. For this tree, however, already $\vert\mathcal{S}\vert=10^4$ yields a ZPVE within \SI{0.2}{cm^{-1}} of the exact value, compared with \SI{3.0}{cm^{-1}} for the nhCDVR.
Over the entire calculation, the nhCDVR required a similar total number of PES evaluations as SDVR with $\vert\mathcal{S}\vert=10^4$.
With a different tree design or the addition of extra points to improve the quadrature, the nhCDVR yields the correct ZPVE\cite{Ellerbrock2022,Ellerbrock2023,Manthe2026}.
Owing to its construction, the nhCDVR does not support the state-by-state eigensolver with the penalty of Eq.~\eqref{eq:penalty} employed here, and excited states have to be computed with a state-averaged solver.
A comparison of excited states would mix the effects of the potential representation and of the eigensolver, and is therefore not presented.

\subsection{Eigenstates of Acetonitrile}
\label{sec:mecn}

\begin{figure*}[t]
\includegraphics{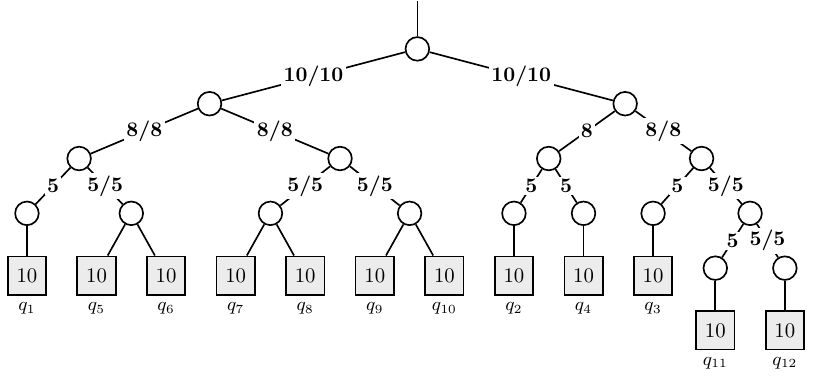}
\caption{TTNS topology employed in the calculation of the vibrational eigenstates of acetonitrile.
The tensor network is symmetry-adapted.
Where two numbers are given, they denote the bond dimensions of the $A'$ and $A''$ sectors.
A single number means that only the $A'$ sector is present.}
\label{fig:mecn-tree}
\end{figure*}

The vibrational states of acetonitrile~(12D) have served as a benchmark for many methods\cite{Avila2011,Leclerc2014,Halverson2015,Thomas2015,Rakhuba2016,Wodraszka2016,Avila2017,Baiardi2017,Odunlami2017,Larsson2019,Wodraszka2020,Fetherolf2021,Sarka2021,Simmons2023,Kallullathil2023,Hoppe2024,Zhang2024,Larsson2025}.
In the context of this work, the system is of particular interest because it requires a more complex TTNS topology, shown in Fig.~\ref{fig:mecn-tree}.
The topology is taken from Ref.~\onlinecite{Larsson2019}, and the bond dimensions were chosen heuristically.
The PES is the quartic force field of B\'egu\'e~\textit{et al.}\cite{Begue2005} in dimensionless normal coordinates $q_k$.
The kinetic energy operator is $\hat T=-\frac{1}{2}\sum_k\omega_k\partial^2/\partial q_k^2$, where $\omega_k$ denotes the harmonic frequency of mode $k$.
For each mode, the primitive basis is a harmonic-oscillator DVR with unit frequency in $q_k$, with the grid sizes given in Fig.~\ref{fig:mecn-tree}.
The exact reference uses the SOP form of this PES in the same DVR.
SDVR instead evaluates the PES pointwise, as for a general PES.
The TTNS are symmetry-adapted following Singh~\textit{et al.}\cite{Singh2010} and Gunst~\textit{et al.}\cite{gunst2019}.
Each node stores one tensor per irreducible representation of the point group, and the full coefficient tensor is their direct sum.
The $C_s$ subgroup of the molecular point group $C_{3v}$ is used, and the bond dimensions in Fig.~\ref{fig:mecn-tree} are given per irreducible representation.
If only one number is given, only the $A'$ sector is present.
In total, 25 vibrational eigenstates were calculated, 15 of $A'$ and 10 of $A''$ symmetry, all with $\mu=0.25$.
The resulting energies are given in Table~\ref{tab:mecn_vmc}.

\begin{table}
\centering
\caption[Vibrational energies of acetonitrile.]{Zero-point energy and vibrational excitation energies of acetonitrile in \si{cm^{-1}}.
The exact results are obtained by using the SOP form of the Hamiltonian.
All other results are obtained with the given $\vert\mathcal{S}\vert$.
Differences $\Delta E = E - E_\text{exact}$ indicate the deviation from the exact result.
Note that excitation energies and not absolute energies are compared (except for the ZPVE).
All excitation energies are relative to the $A'$ ground state. MAE and RMS are computed over the excitation energies only.}
\label{tab:mecn_vmc}
\footnotesize
\setlength{\tabcolsep}{4pt}
\begin{tabular}{rcr|rr}
\toprule
 & & exact & \multicolumn{2}{c}{\begin{tabular}{@{}c@{}}$\vert\mathcal{S}\vert = 10^5$\end{tabular}} \\\hline
\cmidrule(lr){4-5}
State & Sym. & $E$ & $E$ & $\Delta E$ \\
\midrule
1 & $A'$ & 9837.4 & 9837.4 & 0.0 \\
\midrule
2 & $A''$ & 361.0 & 361.0 & 0.0 \\
3 & $A'$ & 361.0 & 361.1 & 0.1 \\
4 & $A''$ & 723.2 & 723.3 & 0.1 \\
5 & $A'$ & 723.2 & 723.3 & 0.1 \\\hline
6 & $A'$ & 723.8 & 723.8 & 0.0 \\
7 & $A'$ & 900.7 & 900.7 & 0.0 \\
8 & $A''$ & 1034.1 & 1034.2 & 0.1 \\
9 & $A'$ & 1034.2 & 1034.1 & -0.1 \\
10 & $A''$ & 1086.6 & 1086.6 & 0.0 \\\hline
11 & $A'$ & 1086.6 & 1086.7 & 0.1 \\
12 & $A''$ & 1087.8 & 1087.8 & 0.0 \\
13 & $A'$ & 1087.8 & 1087.8 & 0.0 \\
14 & $A''$ & 1259.8 & 1259.9 & 0.1 \\
15 & $A'$ & 1259.8 & 1259.9 & 0.1 \\\hline
16 & $A'$ & 1389.3 & 1389.4 & 0.1 \\
17 & $A''$ & 1394.8 & 1394.8 & 0.0 \\
18 & $A''$ & 1394.8 & 1394.9 & 0.1 \\
19 & $A'$ & 1395.0 & 1395.3 & 0.3 \\
20 & $A'$ & 1398.0 & 1398.3 & 0.3 \\\hline
21 & $A''$ & 1451.1 & 1451.1 & 0.0 \\
22 & $A'$ & 1451.2 & 1451.3 & 0.1 \\
23 & $A''$ & 1452.9 & 1452.9 & 0.0 \\
24 & $A'$ & 1453.1 & 1452.9 & -0.2 \\
25 & $A'$ & 1453.6 & 1453.6 & 0.0 \\\hline
\midrule
MAE &  &  &  & 0.1 \\
RMS &  &  &  & 0.1 \\
\bottomrule
\end{tabular}
\end{table}

The primitive grid spans a total of $10^{12}$ configurations, and only $10^5$ are sampled per matrix update.
Still, the excitation energies agree with the exact results with an MAE of about \SI{0.1}{cm^{-1}}.
As for CH$_3$, most nonzero deviations are positive.
This is consistent with the bias of the optimized eigenvectors.
The largest deviations, \SIrange{0.2}{0.3}{cm^{-1}}, occur for states 19, 20, and 24.
These lie in the dense region between 1395 and \SI{1454}{cm^{-1}}, where several states lie within a few wavenumbers of each other.
The calculation required about twice as many sweeps per eigenstate as the exact reference calculation, it converges quickly despite the sampling noise.
On average, each eigenstate required 8 sweeps of 33 node updates each.
For all 25 states, this amounts to $8\cdot25\cdot33\cdot\vert\mathcal{S}\vert\approx6.6\times10^{8}$ PES evaluations, about 0.07\% of the $10^{12}$ grid points.
The calculation took about 0.5 hours on 8 cores, of which about 80\% was spent evaluating the PES.

Overall, the eigenenergies of acetonitrile agree very well with the exact reference.
SDVR thus remains accurate for larger systems with more complex tree topologies.

\subsection{Zero-Point Vibrational Energy of Uracil}
\label{sec:uracil}

\begin{figure*}
    \centering
    \includegraphics[width=0.9\textwidth]{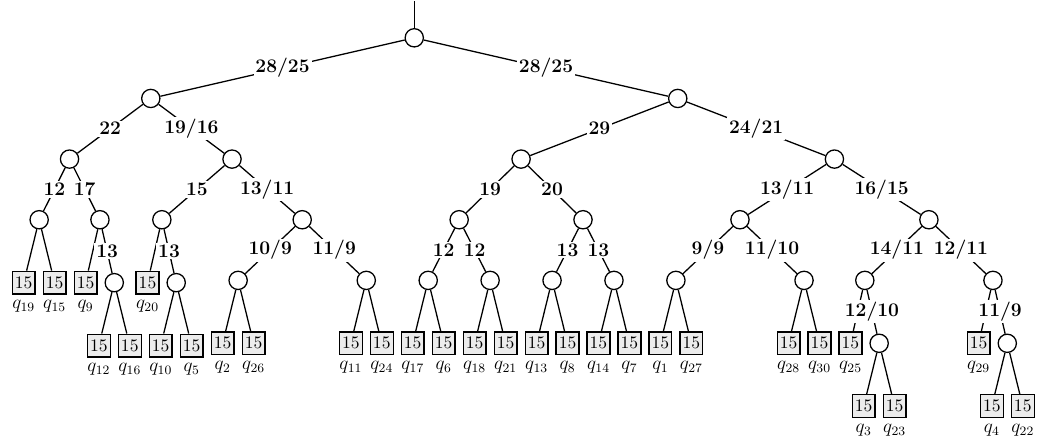}
    \caption{TTNS topology employed in the calculation of the ZPVE of uracil.
    The primitive grid size is 15 for all degrees of freedom.
    The tensor network is symmetry-adapted.
    Where two numbers are given, they denote the bond dimensions of the $A'$ and $A''$ sectors.
    A single number means that only the $A'$ sector is present.}
    \label{fig:uracil}
\end{figure*}

The final test example is the calculation of the ZPVE of the 30-dimensional uracil molecule.
The kinetic energy operator has the same normal-mode form as for acetonitrile (Sec.~\ref{sec:mecn}), and the same harmonic-oscillator DVR is used as primitive basis.
The PES is the quartic force field of Ref.~\onlinecite{thomas2018}. Since the bare quartic force field has holes, its coupling terms are damped by the tanh function introduced there ($\xi=1.5$).
Although this PES is of SOP form, it is evaluated pointwise, as for a general PES.
The TTNS topology is shown in Fig.~\ref{fig:uracil}.
In this topology, the strongly correlated pairs of stretching and out-of-plane modes end up as pairs at the bottom of the tree.
The primitive grid comprises $15^{30}\approx2\times10^{35}$ points.
The calculation employs $\vert\mathcal{S}\vert=10^8$, $\mu=0.95$, and a $2$-mode control variate.
As for acetonitrile (Sec.~\ref{sec:mecn}), a symmetry-adapted TTNS is used, here for the $C_s$ point group.

The calculation converged after 18 sweeps.
Over the last 3 sweeps, the VMC energy changed by less than \SI{0.1}{cm^{-1}}.
It took about 24 hours on a 192-core machine.
With $\vert\mathcal{M}\vert=10^8$, SDVR gives a ZPVE of \SI{18990.44(6)}{cm^{-1}}.
This lies \SI{0.10(6)}{cm^{-1}} above the value of \SI{18990.34}{cm^{-1}} obtained with the same TTNS and the SOP form of the Hamiltonian.
The deviation is not significant, but its sign is consistent with the bias of the optimized eigenvector discussed in Sec.~\ref{sec:noise-control}.
Both values refer to the TTNS of Fig.~\ref{fig:uracil}.
Since the energy is variational with respect to the bond dimensions, larger bond dimensions can only lower the ZPVE further.
About 95\% of the runtime was spent evaluating the PES.
Even with a thousand times more VMC samples than for CH$_3$ and acetonitrile, the statistical error is several times larger, the local energies vary considerably more.
Below the top node, the bond dimensions are large.
For a given $\vert\mathcal{S}\vert$, the sampled operators there are noisier (cp.\ Eq.~\eqref{eq:varbound}).
The optimized state ends further from an eigenstate, and its local energies vary more.
The method can thus in principle reach high accuracy for large systems.
The variance of the sampled matrices, however, can make this numerically expensive.

\section{Conclusions}

A stochastic discrete variable representation~(SDVR) of the potential operator in the context of tensor network states has been presented.
The matrix elements of a general potential operator in the node-local configuration space are estimated by importance sampling of the primitive grid configurations. 
These configurations are drawn exactly and independently by perfect sampling over both SPFs and SHFs of the neighboring branches.
The resulting potential energy matrices are bounded, and only pointwise evaluations of the PES are required.
The variance of the sampled operator is reduced by employing a control variate, e.g., an $n$-mode representation of the potential.
Apart from generating a control variate, no fit of the PES is required.
Noise arising in the eigenproblem from the variance of the sampled operator is mitigated, but not entirely removed, by a simple scheme.
The final energies are not obtained from the eigenproblem, but instead via the unbiased variational Monte Carlo estimator.

For the methyl radical, the $2$-mode representation alone gives a mean absolute error of about \SI{5}{cm^{-1}}.
Sampling as little as $10^4$ configurations per node update reduces it to \SIrange{0.2}{0.3}{cm^{-1}}. 
Employing $10^5$ configurations further reduces that error to about \SI{0.03}{cm^{-1}}.
Acetonitrile requires a deeper, symmetry-adapted tree.
Its 25 vibrational states are reproduced with a mean absolute error of about \SI{0.1}{cm^{-1}}, while evaluating the PES at only $10^5$ of the $10^{12}$ grid points per node update, and at less than 1\% of them over the entire calculation.
For uracil~(30D), the ZPVE of \SI{18990.44(6)}{cm^{-1}} agrees with the SOP result within twice its statistical error.
For the given tree, larger bond dimensions would lower the ZPVE further.
Its local energies vary considerably more than for the smaller systems.
The larger bond dimensions make the sampled operators noisier for a given number of samples.

The accuracy of SDVR can be controlled via two parameters, $\mu$ and the number of samples.
The mixing parameter $\mu$ controls how strongly unoccupied functions on the bonds of the canonical center are sampled.
In the examples studied here, it has only a minor effect, with a non-significant deviation between $\mu=0.25$ and $\mu=0.75$ at small sample sizes.
Increasing the number of samples systematically improves the results, as shown for CH$_3$, and the VMC estimator provides statistical error bars for the final energies.
For small systems, SDVR is not the most economical choice. 
When the PES can be accurately refitted into SOP form, e.g., by the POTFIT family of algorithms, this strongly reduces the numerical effort.
For small molecules, the CDVR with a reasonable tree topology is accurate and cheap.
The main expected advantage of SDVR lies in high-dimensional systems with general PESs, where SOP fits and the CDVR become costly or unreliable.
SDVR only changes how the potential matrix elements are obtained.
It can therefore be combined with any TTNS eigensolver based on the action of the Hamiltonian, such as state-averaged solvers or inexact Lanczos methods for excited states\cite{Rano2025}.

The variance bound of the sampled operator grows with the product of the bond dimensions at a node.
The bias of the optimized eigenvectors decreases with the number of samples. Future work will therefore target variance reduction. Promising options are more accurate control variates, for which any SOP fit of the PES can serve, and node-dependent sample sizes.

\section{Acknowledgments}
HH was supported by an ETH Zurich Postdoctoral Fellowship (25-2 FEL-020). 
Claude (Anthropic) was used to support parts of the implementation of numerical algorithms and final touches to the manuscript.
The author thanks (in alphabetical order) Uwe Manthe, Jeremy Richardson, and Yu-Chen Wang~\begin{CJK*}{UTF8}{bsmi}
(汪宇晨)
\end{CJK*} for the helpful discussions throughout this project.

\section*{Author Declarations}
The author has no conflicts to disclose.

\section*{Data Availability}
The data that support the findings of this study are available within the article.

\bibliography{references}

\end{document}